\documentclass[epj]{webofc}
\usepackage[utf8]{inputenc}
\usepackage[varg]{txfonts}   
\usepackage{booktabs}
\usepackage{xcolor}
\definecolor{darkred}{rgb}{0.4,0.0,0.0}
\definecolor{darkgreen}{rgb}{0.0,0.4,0.0}
\definecolor{darkblue}{rgb}{0.0,0.0,0.4}
\usepackage[bookmarks,linktocpage,colorlinks,
    linkcolor = darkred,
    urlcolor  = darkblue,
    citecolor = darkgreen]{hyperref}
\usepackage{subfigure}
\usepackage{dsfont}
\wocname{EPJ Web of Conferences}
\woctitle{Lattice2017}
\newcommand{\eins}{\mathds{1}}
\newcommand{\Ob}{\mathcal{O}}

\newcommand{\fpiK}{f_{\pi K}}
\newcommand{\mR}{\overline{m}}
\newcommand{\ggGF}{\bar g^2_{\text{GF}}}
\newcommand{\ggSF}{\bar g^2_{\text{SF}}}
\newcommand{\gbar}{\bar g}
\newcommand{\Nf}{N_{\rm f}}
\newcommand{\Lhad}{L_{\text{had}}}
\newcommand{\Lms}{\Lambda^{(3)}_{\overline{\text{MS}}}}
\newcommand{\eq}[1]{Eq.~(\ref{#1})}
\newcommand{\fig}[1]{Fig.~\ref{#1}}
\newcommand{\tab}[1]{Tab.~\ref{#1}}
\newcommand{\sect}[1]{Sec.~\ref{#1}}
\begin{document}
%
\selectlanguage{english}
\title{%
Determination of the Strong Coupling Constant by the ALPHA Collaboration
}
\author{%
\firstname{Tomasz} \lastname{Korzec}\inst{1}
\fnsep\thanks{Speaker, \email{korzec@uni-wuppertal.de}}
for the ALPHA collaboration}
\institute{%
Department of Physics, Bergische Universit\"at Wuppertal, Gau\ss str. 20, 42119 Wuppertal, Germany
}
\abstract{%
A high precision determination of the strong coupling constant in the $\overline{\text{MS}}$ scheme at 
the Z-mass scale, using low energy quantities, 
namely pion/kaon decay constants and masses, as experimental input is presented. 
The computation employs two different massless finite volume renormalization schemes to 
non-perturbatively trace the scale dependence of the respective running couplings from a 
scale of about 200 MeV to 100 GeV. At the largest energies perturbation theory is 
reliable. At high energies the Schr\"odinger-Functional scheme is used, while the running 
at low and intermediate energies is computed in a novel renormalization scheme based on an 
improved gradient flow. Large volume $\Nf=2+1$ QCD simulations by CLS are used to set 
the overall scale. The result is compared to world averages by FLAG and the PDG.
}
\maketitle
\section{Introduction}\label{sec-intro}
Parameters of the standard model have to be determined experimentally before
any predictions can be made. Improvements in the knowledge of these
fundamental quantities translate into a higher predictive power
of the model and are crucial for the successful operation of expensive
collider experiments like the LHC. A parameter that is most important
and so far not particularly precisely determined is the coupling in the strong 
sector of the standard model. The difficulties in its determination 
lie in the confining properties 
of QCD. The typical method to measure it is to ``fit'' the perturbative prediction 
of a high energy process to the corresponding measurement. In order for the 
truncated perturbative series to describe the process well, one is forced to 
concentrate on processes where the physical energy scale of the process 
is high,
i.e. $\mu \gg 1$~GeV. Only then the QCD coupling is small enough and truncation
errors are under control. The main uncertainties are systematic errors associated
with the necessary processing of the raw data, before it can be compared to
perturbation theory. To infer what fundamental process has taken place from the 
measured energies and momenta of photons, leptons and hadrons, one relies not only on a
detailed mathematical model of the detector, but also on a good understanding of the
hadronization process. The latter is too complicated to be computed within the 
standard model and various model assumptions enter which in the end may dominate the 
systematic error. Furthermore many experiments rely on a set of measured
structure functions, which can introduce subtle correlations in the results of
different collaborations.

\begin{figure}[thb]
  \centering
  \begin{center}
   \includegraphics[height=6cm]{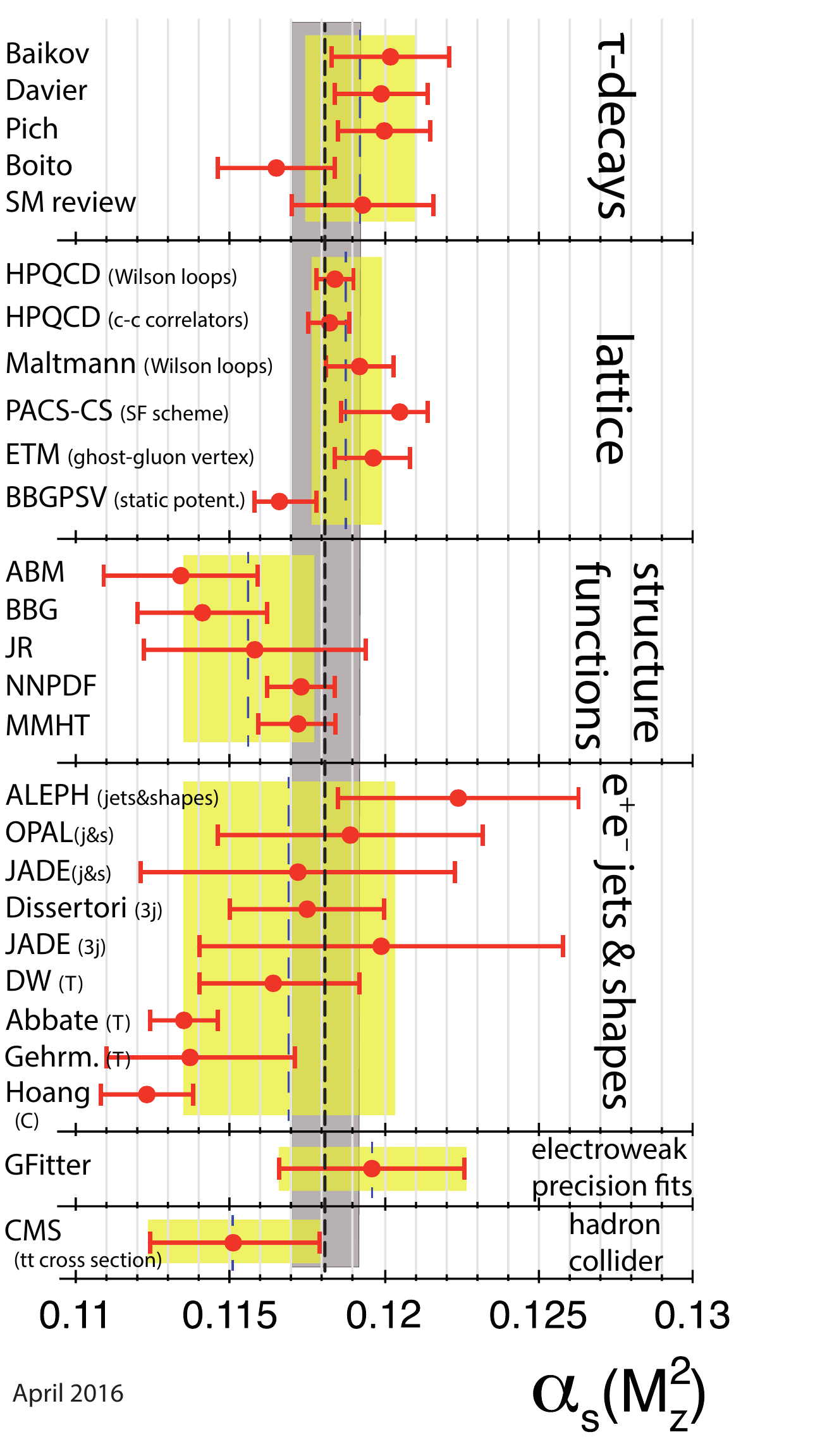}
   \hspace{1cm}
   \includegraphics[height=6cm]{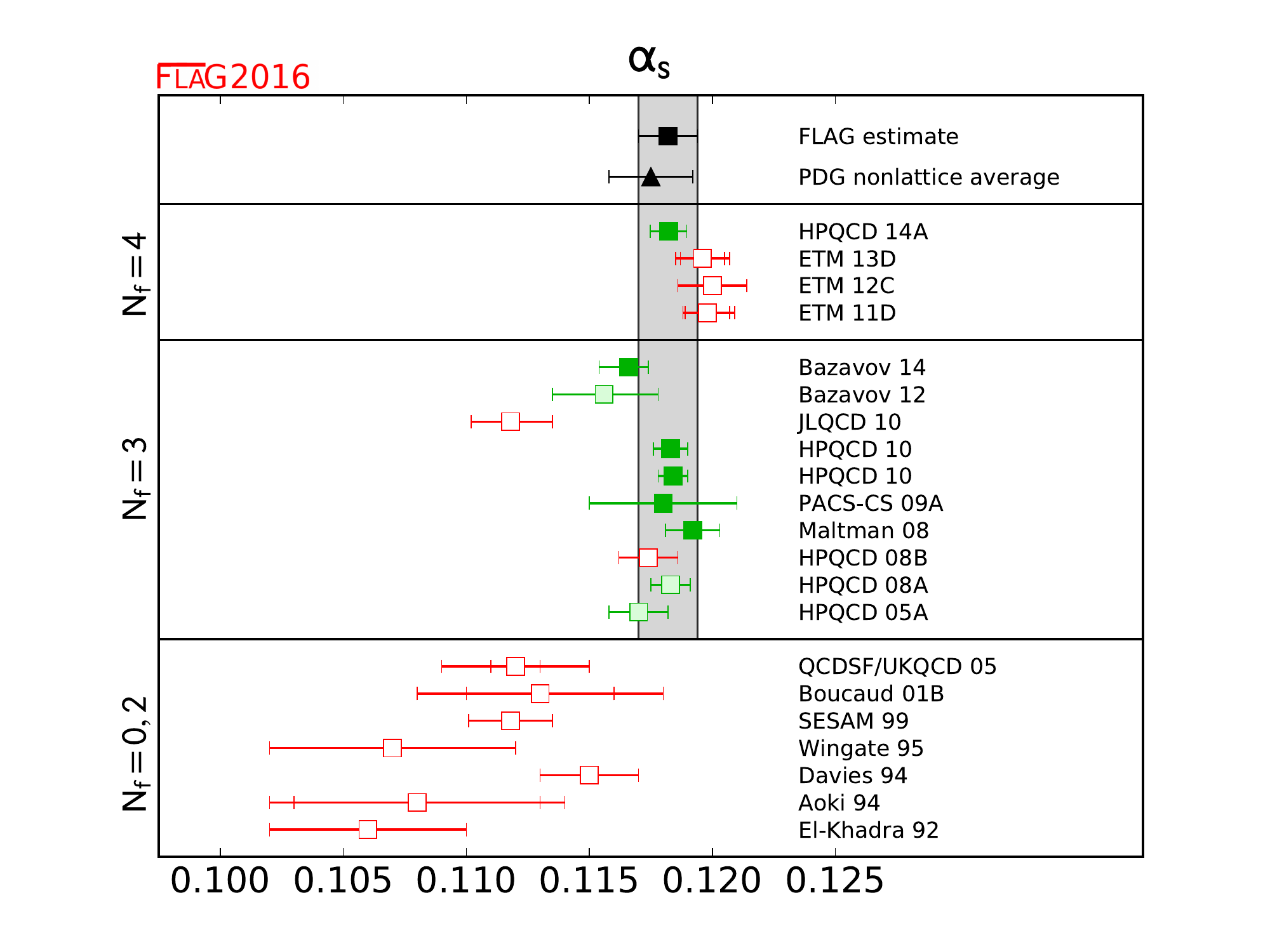}
  \end{center}
  \caption{The left panel is PDG's~\cite{Patrignani:2016xqp} summary of $\alpha_\text{S}$ determinations. The 
  yellow bands correspond to the pre-averages of different classes of determinations.
  The right panel, by
  FLAG~\cite{Aoki:2016frl}, focuses on determinations that used lattice QCD as a tool.
  The gray bands
  are the global averages of the two groups.}\label{fig-world}
\end{figure}

A summary of the methods and most precise experimental determinations is compiled
regularly by the particle data group~\cite{Patrignani:2016xqp}. The different 
results are combined into a world average, its most recent value being
$\alpha_{\rm S}(M_{\rm Z}) = 0.1181(11)$. This is the $\overline{\rm MS}$-coupling
of the five flavor theory renormalized at $\mu=M_{\rm{Z}}$, 
$\alpha_\text{S}(\mu)\equiv \left(\bar g^{(5)}_{\overline{\text{MS}}}(\mu)\right)^2/(4\pi)$. 
In recent years
lattice QCD has become increasingly important as a tool that allows to connect
low and high energy regimes non-perturbatively. The coupling
$\alpha_{\rm S}(M_{\rm Z})$ can then be determined from measured values of 
low energy hadronic quantities
like pion masses and decay constants. Currently some of the world's most
precise determinations are based on lattice QCD and already dominate the world average.
The latest status is summarized in \fig{fig-world}. Lattice determinations are 
also reviewed and averaged by the Flavour Lattice Averaging Group~\cite{Aoki:2016frl}.
The current FLAG average includes results 
from~\cite{Maltman:2008bx,Aoki:2009tf,McNeile:2010ji,Bazavov:2014soa,Chakraborty:2014aca}.
Most of the uncertainties in these lattice determinations are systematic in 
nature and are rooted in the multi-scale nature of the problem. On the one hand
the spatial extent $L$ of the simulated box needs to be large enough to avoid 
finite volume effects in hadronic observables, i.e. $L \gg m_{\pi}^{-1}$.
On the other hand the lattice spacing $a$ must be fine enough to be able
to compute a renormalized coupling at high energies $\mu$ (where it is small).
A coupling could for instance be defined through the static force at short distances $r=\mu^{-1}$,
which would require $a\ll\mu^{-1}$. Insisting on a high value of $\mu$, e.g. $\mu \approx 100$~GeV
immediately leads to astronomically large lattice sizes $L/a$. With today's 
machines and algorithms one is restricted to $L/a \lesssim 100$ which requires a
careful balance of the scales $a$, $\mu$ and $L$ such that the unavoidable
finite size-, cutoff- and perturbative truncation errors are all under control.
For instance the most recent result~\cite{Chakraborty:2014aca} of \fig{fig-world} uses lattices 
with up to $L/a=64$ sites, renormalization scales $\mu\approx 5$~GeV with 
lattice spacings $a^{-1} \approx 3.3$~ GeV and coarser.

These systematic errors can be nearly eliminated by switching to finite volume
renormalization schemes~\cite{Luscher:1991wu}, where $\mu \equiv L^{-1}$. The drawback is,
that a whole sequence of simulations at various values of $L$ becomes necessary.
In addition, the results based on finite size scaling were so far plagued by 
relatively large statistical errors or an insufficient number of dynamical
flavors. The only calculation of this type
that enters the FLAG average is the one by PACS-CS, \cite{Aoki:2009tf}. It has a large, but mainly 
statistical error.

This article summarizes the effort of the ALPHA collaboration to determine 
$\alpha_\text{S}$ using finite-size-scaling techniques. It is self-contained but, due to
size restrictions, many details are omitted. The reader is encouraged to consult the 
original literature~\cite{Bruno:2014jqa,Bruno:2016plf,Brida:2016flw,DallaBrida:2016kgh,Bruno:2017gxd} 
for a deeper understanding of the calculation.

The outline of this proceeding contribution is as follows: in \sect{sec-schemes}
our notation is fixed and basic formulae collected. \sect{sec-strat} lays out 
our computational strategy and the different more or less independent parts of the
calculation are treated in some detail in sections \ref{sec-scale} - \ref{sec-dec}.
The article concludes  with \sect{sec-res} where the final result is put together and
discussed.

\section{Renormalization Schemes and Scales}\label{sec-schemes}
Almost any renormalized, dimensionless, short-distance quantity 
$\phi$ can be the starting point for the definition of a
renormalization scheme. If it possess a perturbative expansion in a bare
coupling $g_0$, e.g. $\phi = \sum\limits_{l} \phi_l g_0^{2l}$, one can define a
renormalized coupling in this scheme as $\gbar_\phi^2 \equiv \frac{\phi-\phi_0}{\phi_1}$.
The quantity $\phi$, the coefficients $\phi_l$ and therefore the renormalized coupling
depend on a scale $\mu$. The scale dependence of the coupling is expressed by the RG equation
\begin{equation}
   \mu \frac{d\gbar}{d\mu} = \beta(\gbar) \, .
\end{equation}
The $\beta$ function depends on the scheme, but in massless schemes the first two 
coefficients in a perturbative expansion
\begin{equation}\label{eq-beta}
   \beta(\gbar) = -\gbar^3\left(b_0 + b_1 \gbar^2 + b_2 \gbar^4 + \ldots \right)
\end{equation}
are ``universal'' 
\begin{equation}
   b_0 = \frac{1}{16\pi^2}\left(11-\frac{2\Nf}{3} \right)\, , \qquad 
   b_1 = \frac{1}{256\pi^4}\left(102 - \frac{38\Nf}{3} \right)\, . 
\end{equation}
The subsequent coefficients depend on the scheme. They are known
to five loops ($b_2$ through $b_4$) in the 
$\overline{\rm MS}$-scheme~\cite{vanRitbergen:1997va,Czakon:2004bu,Baikov:2016tgj,Luthe:2016ima,Herzog:2017ohr}
and to three loops ($b_2$) in the SF-scheme~\cite{Bode:1999sm}.
The solution of the 
ODE~\eq{eq-beta} requires an initial condition, or equivalently, the introduction of a $\Lambda$-parameter, 
e.g. at one loop\footnote{\eq{eq-L1loop} is more of pedagogical than practical value, since it cannot be used to 
determine $\Lambda$, no matter how high $\mu$ is. It does not approximate \eq{eq-Lambda} when $\bar g \to 0$. 
At least $\beta^{2-\text{loop}}$ is needed to properly define $\Lambda$.}
\begin{equation}\label{eq-L1loop}
   \gbar^2(\mu)    = \frac{1}{2 b_0 \ln \left(\mu/\Lambda \right)} \qquad \text{or} \qquad
   \Lambda         = \mu \ e^{-1/(2b_0\bar g^2(\mu))} \, .
\end{equation}
The well known fully non-perturbative expression is
\begin{equation}\label{eq-Lambda}
\Lambda = \mu \left(b_0\gbar^2(\mu) \right)^{-\frac{b_1}{2b_0^2}}\, \exp\left[-\frac{1}{2b_0\gbar^2(\mu)}
          -\int\limits_0^{\gbar(\mu)}\left(\frac{1}{\beta(x)}+\frac{1}{b_0 x^3}- \frac{b_1}{b_0^2 x} \right)dx\right]\ .
\end{equation}
$\Lambda$ parameters are scheme dependent. Between two schemes with couplings $\gbar$ and $\gbar'$ that 
are related in perturbation theory by
\begin{equation}
   \gbar'^2(\mu) = \gbar^2(\mu) + c \gbar^4(\mu) + \ldots \, ,
\end{equation}
the exact relation between $\Lambda$ parameters is
\begin{equation}
   \Lambda' = \Lambda\ e^{c/(2b_0)} \, .
\end{equation}

Instead of the $\beta$-function, the step-scaling function $\sigma(\gbar^2)$ can be invoked to
express the scale dependence of a renormalized coupling. It is defined as the value of the coupling
at a renormalization scale that is smaller by a factor of two
\begin{equation}
   \sigma(u) \equiv \gbar^2(\mu/2)\bigr|_{\gbar^2(\mu)=u}\, .
\end{equation}
Both are related by
\begin{equation}\label{eq-betatossf}
   -\ln 2 = \int\limits_{\sqrt{u}}^{\sqrt{\sigma(u)} }\frac{dx}{\beta(x)}  \, .
\end{equation}
The step-scaling function is more directly accessible through computer simulations and plays a 
central role in our calculation. Its perturbative expansion is
\begin{eqnarray}
   \sigma(u) &=& u + s_0 u^2 + s_1 u^3 + s_2 u^4 + \ldots \\
   s_0       &=& 2b_0 \ln 2\, , \qquad
   s_1      \ = \ (2b_0 \ln 2)^2 + 2b_1 \ln 2\, , \qquad \text{etc.}
\end{eqnarray}

\section{Strategy}\label{sec-strat}
Lattice QCD is used as a non-perturbative tool that allows to determine the $\Lambda$ parameter of QCD
from low energy experiments.

Our computational strategy can be summarized as follows. In large volume QCD simulations the value of a hadronic quantity,
like the pion or kaon decay constant (or, in our case, a linear combination of both, denoted by $f_{\pi K}$), 
is determined at the physical mass point in lattice units on lattices with
different lattice spacings. The experimental measurement of the same quantity\footnote{In pure QCD $f_\pi$
is defined by a matrix element of the axial current between the vacuum and the pion state.
Relating this to the experimentally measured values is intricate~\cite{Gasser:2010wz}
and to some extent convention dependent.
} allows to map out a relationship
between the bare coupling $g_0$ of the chosen lattice action and the lattice spacing in fm. At this point
one can switch to massless small volume simulations with a known box size $\Lhad$. A 
renormalized coupling $\gbar(\mu)$ in a finite volume scheme is defined, in which the renormalization 
scale is tied to the linear box size
$\mu = L^{-1}$. The $\beta$ function of this coupling is determined non-perturbatively for a wide range 
of couplings, by simulations of 
lattices with decreasing box sizes. Once the $\beta$ function is known, it is used to determine the value
of the renormalized coupling at a high energy scale $\mu=\mu_{\rm PT}\approx 70$ GeV, provided that its value 
at $\mu\equiv\Lhad^{-1}$
is known. At these high energies the coupling is small and perturbation theory becomes reliable and is
used to relate the coupling to the $\Lambda$ parameter, i.e. a perturbative approximation of $\beta$ is
used in \eq{eq-Lambda}.
This can then be translated exactly into 
the $\Lambda$ parameter of a more common scheme. All of this is carried out in massive 
($\Nf=2+1$, large volume) or
massless ($\Nf=3$, finite volume) QCD. To obtain the $\Lambda$ parameter in four, five or six flavor QCD,
perturbative decoupling~\cite{Weinberg:1980wa,Bernreuther:1981sg,Grozin:2011nk,Chetyrkin:2005ia,Schroder:2005hy,Kniehl:2006bg}
is invoked. 

In practice the above procedure is extended by two steps. The first is to use two different 
finite volume schemes, each
offering significant advantages in the energy region where they are utilized. The couplings of these two
schemes are matched non-perturbatively at an intermediate scale $L_0^{-1}\approx 5$~GeV. The other step
is to use the flow scale $t_0$~\cite{Luscher:2010iy} as an intermediate scale during the scale-setting, i.e. to first determine
the dimensionless combination of $t_0$ and $\fpiK$ and then the ratio $\Lhad/\sqrt{t_0}$.
In total  the calculation can be summarized by
\begin{equation}\label{eq-Lambdastrat}
  \Lambda_{\overline{\rm MS}}^{(3)} = \underbrace{\frac{f_{\pi K}^{\text{PDG}}}{f_{\pi K}\sqrt{t_0}}}_{\substack{\text{scale setting}\\ \text{\sect{sec-scale}}}}
         \times \underbrace{\frac{\sqrt{t_0}}{L_{\text{had}}}}_{\substack{\text{connection to }L=\infty\\ \text{\sect{sec-Linf}}}} 
         \times \underbrace{\frac{L_{\text{had}}}{2L_{\text{0}}}}_{\substack{\text{scheme 1}\\ \text{\sect{sec-GF}} }}
         \times \underbrace{\frac{2 L_{\text{0}}}{L_\text{0}}}_{\substack{\text{change of schemes}\\ \text{\sect{sec-Lswi}}}}
         \times \underbrace{\Lambda_{\overline{\rm MS}}^{(3)}L_{\text{0}}}_{\substack{\text{scheme 2}\\ \text{\sect{sec-SF}}}} \, ,
\end{equation}
where each factor is largely independent from the others. 

\section{Scale Setting}\label{sec-scale}
The experimental inputs needed for the determination of the $\Lambda-$parameter enter the 
computation through the process of scale-setting. I.e. the relation between the bare coupling
$g_0$ and the lattice spacing in fm depends on these inputs, for which we take
\begin{eqnarray}
   \fpiK^{\rm phys} \equiv \frac{2}{3}f_K^{\rm phys} + \frac{1}{3}f_\pi^{\rm phys} &=& 147.6(5)\ \text{MeV}\, ,  \\
   m_\pi^{\rm phys} &=& 134.8(3)\ \text{MeV}\, ,\\
   m_K^{\rm phys}   &=& 494.2(3)\ \text{MeV} \, .
\end{eqnarray}
The meson masses, corrected for isospin breaking and electro-magnetic effects, are taken
from FLAG~\cite{Aoki:2016frl}, the decay constants from the PDG~\cite{Agashe:2014kda}.

The large-volume simulations were carried out within the ``Coordinated Lattice Simulations'' 
consortium (CLS)~\cite{Bruno:2014jqa}. The set of ensembles was generated using a tree-level Symanzik
improved gauge action~\cite{Luscher:1984xn} and 2+1 flavors of non-perturbatively clover 
improved~\cite{Sheikholeslami:1985ij,Bulava:2013cta} Wilson fermions. Open boundary conditions in temporal 
directions made simulations at
small lattice spacings, down to $a\approx 0.039$ fm possible, without facing problems
due to topological critical slowing down~\cite{Luscher:2011kk,Schaefer:2010hu}. All simulations 
were carried out using the \verb+openQCD+ simulation suite\footnote{\url{http://luscher.web.cern.ch/luscher/openQCD/}}
\cite{Luscher:2012av}.

Finite lattice spacings and unphysical quark masses make a chiral-continuum extrapolation necessary
before the scale can be set.
It is convenient to use the flow scale $t_0$~\cite{Luscher:2010iy} as an intermediate scale
and to measure
\begin{eqnarray}
   8 t_0 m_\pi^ 2&\equiv&\phi_2 \hspace{2cm}  \sim m_\text{up}\, ,\label{eq-phi2}\\
   8t_0\left(m_K^2 + \frac{m_\pi^2}{2} \right)&\equiv&\phi_4 \hspace{2cm}  \sim m_\text{up}+m_\text{down}+m_\text{strange}\, ,\\
   \sqrt{t_0} \fpiK\, ,&& \\
   t_0/a^2 && \label{eq-t0aa}
\end{eqnarray}
on all ensembles. 
For a precise definition of these observables, bare parameters and 
choices of plateau regions, we refer the reader to~\cite{Bruno:2016plf}.
The parameters of the CLS~\cite{Bruno:2014jqa} ensembles were chosen such that
$\phi_4$ is approximately constant and close to its physical value.
Values of $\phi_2$ span a range corresponding to $200$~MeV~$\lesssim m_\pi \lesssim 420$~MeV.
The simulated bare couplings correspond to four different lattice spacings
between 0.04 fm and 0.09 fm. 
For the dependence of $\sqrt{t_0} \fpiK$ on $\phi_2$ and the lattice spacing, different 
assumptions can be made. A Taylor expansion around the flavor-symmetric point
motivates the ansatz~\cite{Bietenholz:2010jr}
\begin{equation}\label{eq-taylor}
   f^{\rm Taylor}(\phi_2,a) = c_0 + c_1 (\phi_2-\phi_2^\text{sym})^2  + c_2 \frac{a^2}{t_0^\text{sym}} \, ,
\end{equation}
while chiral perturbation theory~\cite{Bar:2013ora,Gasser:1984ux} suggests
\begin{equation}\label{eq-chipt}
   f^{\chi{\rm PT}}(\phi_2,a) = (\sqrt{t_0}f_{\pi K})^\text{sym}\
          \left[1-\frac{7(L_\pi-L_\pi^\text{sym})}{6}-\frac{4(L_K-L_K^\text{sym})}{3}-\frac{L_\eta-L_\eta^\text{sym}}{2}  \right]
          +c_4 \frac{a^2}{t_0^\text{sym}} \, ,
\end{equation}
with logarithms  $L_x = \frac{m_x^2}{(4\pi f)^2}\, \ln\left[\frac{m_x^2}{(4\pi f)^2} \right]$.
These and other functions can be used to read off the value at the 
physical point. The intermediate scale $\sqrt{t_0^{\rm phys}}$ in 
fm is given by $\sqrt{t_0^{\rm phys}} = f(\phi_2^{\rm phys},0) / \fpiK^{\rm phys}$.
The lattice spacings in fm then follow from the $t_0/a^2$ measurements, 
extrapolated to $\phi_2=\phi_2^{\text{phys}}$.

\begin{figure}[thb]
  \centering
  \begin{center}
   \includegraphics[width=\linewidth]{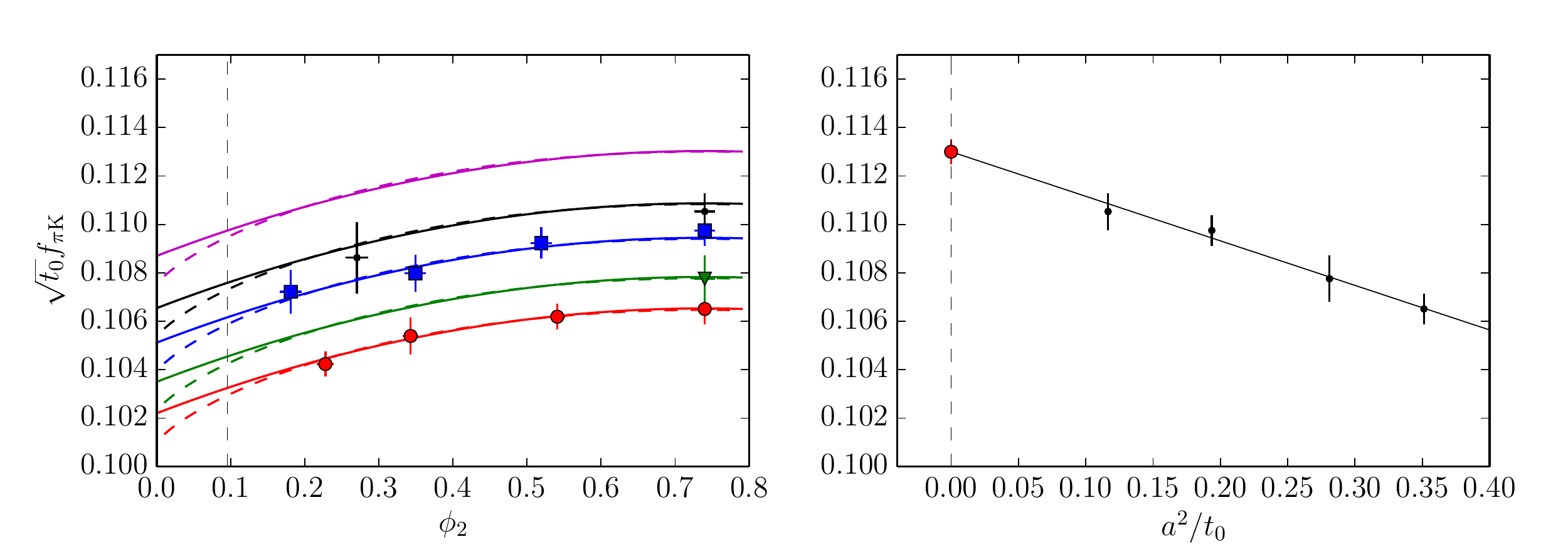}
  \end{center}
  \caption{The left panel shows chiral and continuum extrapolations of $\sqrt{t_0}\fpiK$. Solid lines correspond to 
  extrapolations according to \eq{eq-taylor} and dashed lines to \eq{eq-chipt}. From bottom to top the 
  lattice spacing decreases, the corresponding bare couplings $6/g_0^2$ are 3.4 (red), 3.46 (green), 3.55 (blue)
  and 3.7 (black). The extrapolated continuum curve (magenta) is on top. In the right panel it is shown, how well
  the global fit describes the lattice spacing dependence at the flavor symmetric mass point, where data is
  available for all lattice spacings.}
  \label{fig-chicont}
\end{figure}

The value of $t_0^{\rm phys}$ that was used to plan the simulations and fix the
chiral trajectory was slightly different than the final result of the procedure above.
And even if not, the tuning of the ensembles only has a finite precision.
Small corrections of the mis-tuning and small changes
of the target values are necessary. They can be carried out, if the derivatives 
of all observables \eq{eq-phi2}-\eq{eq-t0aa} with respect to the bare masses are known. 
These derivatives are measured as 
described in~\cite{Bruno:2016plf} and corrections
\begin{equation}
   \Ob(m'_0) = \Ob(m_0) + (m'_0-m_0) \frac{d\Ob}{dm_0} + O\left((m'_0-m_0)^2\right)
\end{equation}
are made until a fixed point in the value of  $t_0^{\rm phys}$ is found.
\fig{fig-chicont} shows two of the chiral-continuum extrapolations that were
attempted after all ensembles were mass-shifted such that $\phi_4=\phi_4^{\rm phys}$.

A further improvement of the procedure above is to replace
$\sqrt{t_0} \fpiK$ by $\sqrt{t_0^\star} \fpiK$, where $t_0^\star$
is defined on the unphysical mass-point where $m_\text{up}=m_\text{down}=m_\text{strange}$ and $\phi_4 = 1.11$.
The advantages are that at this mass-point simulations are comparatively cheap,
no chiral extrapolations of $t_0/a^2$ are necessary, finite volume effects are smaller and, by 
definition, the mass-point remains unchanged, even if in the future, with higher 
statistics, the values of $\phi_2^{\rm phys}$ and $\phi_4^{\rm phys}$ should change.

\begin{table}[thb]
  \parbox{.45\linewidth}{
  \small
  \centering
  \caption{Bare couplings and the corresponding lattice spacings of the CLS ensembles.}\label{tab-t0}
  \begin{tabular}{lll}\toprule
  $6/g_0^2$  & $t_0^\star / a^2$ & $a$ in fm \\\midrule
   3.40 & 2.862(6)   & 0.086 \\
   3.46 & 3.662(13)  & 0.076 \\
   3.55 & 5.166(18)  & 0.064 \\
   3.70 & 8.596(31)  & 0.050 \\
   3.85 &14.036(57)  & 0.039 \\
   \bottomrule
  \end{tabular}
  }
  \hfill
  \parbox{.45\linewidth}{
  \small
  \centering
  \caption{Bare couplings such that $\ggGF(\Lhad^{-1}) = 11.31$ for various $\Lhad/a$.}
  \label{tab-Lhad}
  \begin{tabular}{ll}\toprule
  $6/g_0^2$  & $\Lhad / a$ \\\midrule
   3.3998 & 12.000(58)   \\
   3.5498 & 16.000(30)  \\
   3.6867 & 20.000(83)  \\
   3.8000 & 24.000(105)  \\
   3.9791 & 32.000(153)  \\
   \bottomrule
  \end{tabular}
  }
\end{table}

Based on the CLS ensembles the procedure sketched above leads to
\begin{equation}
   \sqrt{8 t_0^\star} = 0.413(5)(2)\ \text{fm}\, .
\end{equation}
The first error is statistical, the second accounts for uncertainties 
related to the chiral extrapolations. Different functional forms have been tried
and subsets of data neglected/included in order to asses its size~\cite{Bruno:2016plf}.
The resulting lattice spacings are summarized in \tab{tab-t0}.

\section{Running}\label{sec-run}
\begin{figure}[thb]
  \centering
  \begin{center}
   \begin{tabular}{l l c l l}
      $ m_0^{(1)}, g_0^{(1)}$: & $\begin{array}{l}\includegraphics[width= 0.75cm]{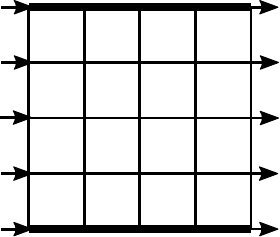} \end{array}$          & $\begin{array}{l}\overset
{\text{same}\ a^{(1)}}{\leftrightarrow}\end{array}$ 
      & $\begin{array}{l}\includegraphics[width= 1.5cm]{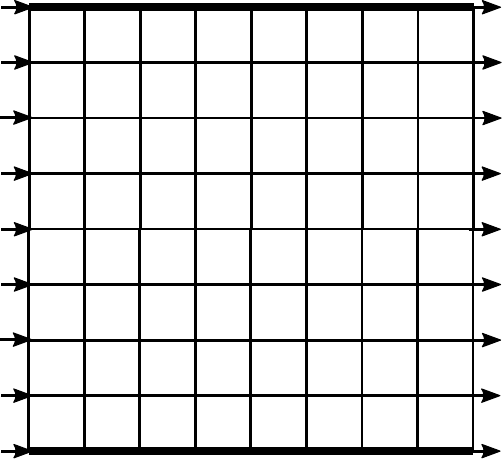}\end{array}$           & $ \to \Sigma(u,a^{(1)}/L)$\\ 
                                     & $\begin{array}{l}\updownarrow$ same $L$, $\bar g^2(L^{-1})\end{array}$            &    &   & \\
      $ m_0^{(2)}, g_0^{(2)}$: & $\begin{array}{l}\includegraphics[width= 0.75cm]{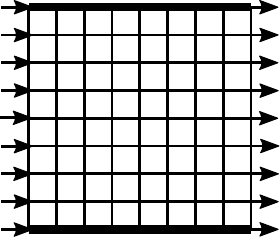}\end{array}$      & $\begin{array}{l}\overset{\text{same}\ a^{(2)}}{\leftrightarrow}\end{array}$ 
      & $\begin{array}{l}\includegraphics[width= 1.5cm]{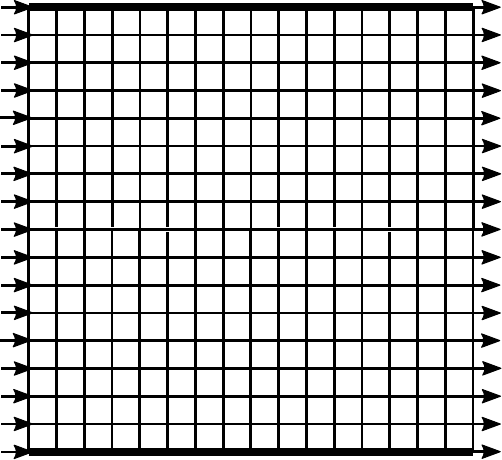}\end{array}$      & $ \to \Sigma(u,a^{(2)}/L)$ \\
                                     & $\begin{array}{l}\updownarrow$ same $L$, $\bar g^2(L^{-1})\end{array}$             &    &   & \\
      $ m_0^{(3)}, g_0^{(3)}$: & $\begin{array}{l}\includegraphics[width= 0.75cm]{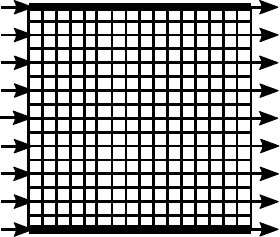}\end{array}$ & $\begin{array}{l}\overset{\text{same}\ a^{(3)}}{\leftrightarrow}\end{array}$ 
      & $\begin{array}{l}\includegraphics[width= 1.5cm]{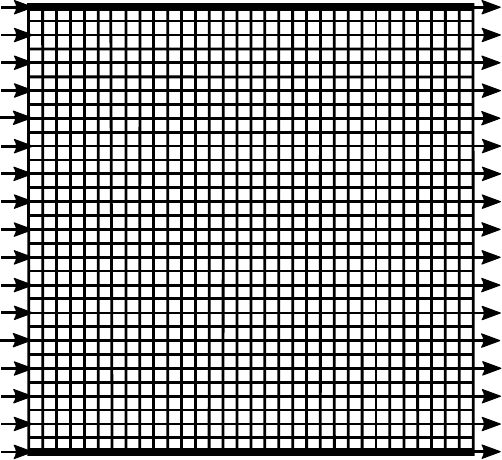}\end{array}$ & $ \to \Sigma(u,a^{(3)}/L)$ \\
                                     &                                                                                     &             
      & $\begin{array}{l}\downarrow $cont. limit $\end{array}$                 & \\
                                     & $\bar g^2=u,\ \mR=0$                                                                           &  
      & $\begin{array}{l}\includegraphics[width= 1.5cm]{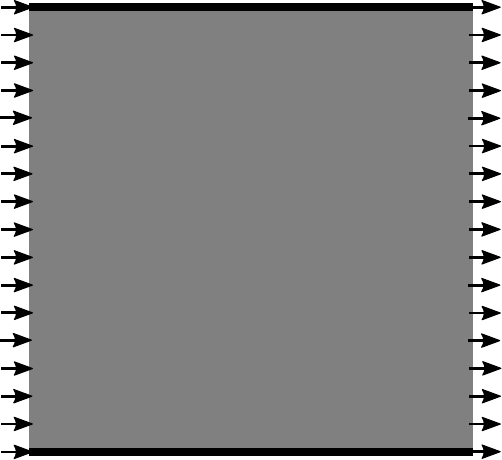}\end{array}$      & $ \to\sigma(u)$ \\
   \end{tabular}
   \end{center}
  \caption{A sketch of the steps necessary to compute one continuum value of a step-scaling function $\sigma$.}
  \label{fig-ssf}
\end{figure}

In finite volume renormalization schemes the renormalization scale is
tied to the linear box size of the world $\mu\equiv L^{-1}$. To determine
the $\beta$ or step-scaling function numerically, a sequence of simulations is
necessary. For instance, to compute $\sigma(u)$ the necessary steps are sketched
in \fig{fig-ssf}. 
First a lattice resolution $L/a$ is chosen. 
Then the bare coupling $g_0$ and the bare mass $m_0$ are tuned such that
renormalized masses are zero, and the renormalized coupling $\gbar^2=u$.
With the same bare parameters another simulation with $2L/a$ lattice points
in each direction is carried out. Up to lattice artifacts, same $g_0$
implies same lattice spacing, so the box-size is doubled (or the renormalization 
scale halved). Measuring the renormalized coupling on the doubled lattice yields 
a lattice estimate of the step-scaling function $\Sigma(u,a/L)$. This has to be
repeated with the same $u$, but finer lattice resolutions, so that a continuum limit
can be taken $\sigma(u) = \lim\limits_{a/L \to 0} \Sigma(u,a/L)$. Finally all of
this has to be repeated at various values of $u$, such that a safe interpolation 
to arbitrary $u$ values becomes possible.

Instead of separate continuum extrapolations for each $u$ and subsequent parametrization of
$\sigma(u)$, both can be combined into one step by performing a global fit. A possible 
ansatz is for instance
\begin{equation}\label{eq-fit1}
   \Sigma(u,a/L) = u + u^2\sum\limits_{k=0}^{n_\sigma}s_ku^k + \left(\frac{a}{L} \right)^2 \sum\limits_{k=1}^{n_\rho}\rho_k u^{k+i+1}
\end{equation}
with the lowest $s_k$ (up to $s_1$ or $s_2$) taken from perturbation theory and the higher left as fit parameters.
Leading lattice artifacts are parametrized by the coefficients $\rho_k$ and
the value of $i$ may be zero, or higher, depending on the order in perturbation theory to which the lattice artifacts were
removed from the raw data. Different combinations of orders of polynomials ($n_\sigma$, $n_\rho$) can be tried.

A somewhat different approach to fit the data is to choose a parametrization of the $\beta$-function e.g.
\begin{eqnarray}
    \beta(g) &=& -g^3\left(b_0 + b_1 g^2 + b_2 g^4 + \ldots \right),  \qquad \text{or} \\
    \beta(g) &=& -\frac{g^3}{p_0 + p_1 g^2 + p_2 g^4 +\ldots}
\end{eqnarray}
and insert it into \eq{eq-betatossf}. In particular the second choice is interesting, because it 
leads to a fit that is linear in the fit parameters $p_k$~\cite{DallaBrida:2016kgh}. Making again
an ansatz for the leading lattice artifacts the fit would minimize the violations
of the equation
\begin{eqnarray}
   \ln 2 &=& -\frac{p_0}{2}\left[\frac{1}{\Sigma(u,a/L)}-\frac{1}{u} \right] + \frac{p_1}{2}\ln\left[\frac{\Sigma(u,a/L)}{u} \right]
             + \sum\limits_{n=1}^{n_p} \frac{p_{n+1}}{2n}\left[\Sigma^n(u,a/L)-u^n \right] \nonumber \\
         &&  + \left(\frac{a}{L}\right)^2 \sum\limits_{k=1}^{n_\rho}\rho_k u^{k+i+1} \, ,
\end{eqnarray}
where $\Sigma(u,a/L)$ are the data.

The first choice leads to a non-linear fit, which is feasible if one
can get away with very few parameters $b_3, b_4,\ldots $ and 
take $b_0,b_1,b_2$ from perturbation theory.

\subsection{QCD Schr\"odinger-Functional}
Boundary conditions play a crucial role for finite volume schemes. Depending on the 
choice simulations with massless fermions become more or less expensive, or not 
possible at all. Schr\"odinger Functional (SF) boundaries are particularly interesting.
They induce a solid spectral gap in the Dirac operator~\cite{Sint:1995ch}, allow to define useful
boundary-to-bulk and boundary-to-boundary correlation functions~\cite{Jansen:1995ck,Capitani:1998mq}
and can be used to
define finite volume couplings with very good properties.

SF boundaries consist of Dirichlet boundaries in time for both fermion and gauge fields, and
periodic (up to a phase) boundaries in spacial directions. Gauge fields satisfy
\begin{equation}
      U_k(x)|_{x_0=0}   = \exp(a\, C_k)\, , \qquad
      U_k(x)|_{x_0=T}   = \exp(a\, C'_k)\, , \qquad
      U_\mu(x+L\hat k)  = U_\mu(x)\, . 
\end{equation}
For the choice of boundary gauge fields we restrict ourselves to Abelian SU(3) matrices
parametrized by the angles $\eta$ and $\nu$~\cite{Luscher:1992an}\, ,
\begin{equation}
              C_k = \frac{i}{L} 
              \begin{pmatrix} 
                 \eta-\frac{\pi}{3}&                                  & \\
                                   &\eta \nu-\frac{\eta}{2}           & \\
                                   &                                  &-\eta\nu-\frac{\eta}{2}+\frac{\pi}{3}
              \end{pmatrix}, \quad
              C'_k = \frac{i}{L}
              \begin{pmatrix}
                -\eta-\pi          &                                     & \\
                                   &\eta\nu+\frac{\eta}{2}+\frac{\pi}{3} & \\
                                   &                                               & \frac{\eta}{2}-\eta\nu+\frac{2\pi}{3}
              \end{pmatrix}
\end{equation}
For matter fields conditions
   \begin{equation}
       \frac{1}{2}[\eins-\gamma_0]\, \psi(x)|_{x_0=0} = 0\, ,\qquad
       \frac{1}{2}[\eins+\gamma_0]\, \psi(x)|_{x_0=T} = 0\, ,\qquad
       \psi(x+L\hat k)                                = e^{i\theta} \psi(x)\, ,
   \end{equation}
and similarly for $\bar\psi$ are imposed~\cite{Sint:1993un}.

\subsection{Schr\"odinger-Functional Coupling}\label{sec-SF}
The standard definition of a SF-coupling~\cite{Luscher:1992an,Luscher:1993gh} is based on the sensitivity of the 
effective action
\begin{equation}
   \Gamma = \ln\left[ \int \! D[U,\bar\psi,\psi] e^{-S[U,\bar\psi,\psi]}\right]
\end{equation}
to the change of boundaries
\begin{equation}\label{eq-gSF}
   \gbar_{\nu} \equiv k\left[\frac{\partial \Gamma}{\partial \eta}\biggr|_{\eta=0}\right]^{-1}\, .
\end{equation}
This is in fact a whole family of renormalized couplings, the most frequently used one being
$\gbar_{\rm SF} \equiv \gbar_{\nu=0}$. The normalization $k$ is chosen such that
$\gbar^2_{\rm SF} = g^2_0 + O(g_0^4)$,~\cite{Luscher:1993gh}.

The advantages of coupling \eq{eq-gSF} are a good statistical precision, remarkably small lattice artifacts,
relatively small computational costs and a good theoretical understanding~\cite{Sint:1995ch,Bode:1999sm}. 
Its $\beta$ function is known
to three loops. In consequence it has been successfully used for the last quarter century to compute 
$\Lambda-$parameters of $\Nf=0$~\cite{Luscher:1993gh}, $\Nf=2$~\cite{Fritzsch:2012wq}, 
$\Nf=3$~\cite{Aoki:2009tf} and $\Nf=4$~\cite{Tekin:2010mm} QCD and in 
various beyond-the-standard-model applications, mainly related to a composite 
Higgs, see e.g.~\cite{Pica:2017gcb} and references therein.

The main disadvantage is that the statistical precision deteriorates at low energies and
close to the continuum limit. To leading order, the relative 
error of $\ggSF$ is proportional to $\ggSF$,
which makes this scheme particularly useful at higher energies, but problematic when 
$\ggSF$ is large. Moreover the variance of $\frac{\partial \Gamma}{\partial \eta}$
is not a renormalized quantity. It diverges in the continuum limit~\cite{deDivitiis:1994yz}, which 
further increases the 
costs of the already expensive fine lattices. Here, the SF scheme is used at  high
energies $\gtrsim 5$~GeV only.

The step-scaling function of the SF coupling is computed as described in \sect{sec-run} for couplings
$u \in \{1.1089, 1.1845, 1.2657, 1.3627, 1.4808, 1.6173, 1.7943, 2.012\}$, 
on lattices with $L/a \in \{4, 6, 8, 12\}$ (and the corresponding doubled lattices).
$L/a=4$ lattices were excluded from the final analysis, $L/a=12$ lattices are available only for a subset of 
the couplings. The largest coupling implicitly defines the scale $L_0$ at which finite volume schemes
are switched
\begin{equation}\label{eq-L0def}
   \ggSF(L_0^{-1}) \equiv 2.012\, .
\end{equation}
To be able to make most use of available perturbative results, the lattice action for this part of
the project is Wilson's plaquette action with clover-improved Wilson fermions. For this setup 
the coefficient of the clover term is known non-perturbatively~\cite{Yamada:2004ja} 
and the boundary improvement
coefficients $c_t$~\cite{Luscher:1992an}  and $\tilde c_t$~\cite{Luscher:1996sc} 
are known to two~\cite{Bode:1999sm} 
and one~\cite{Luscher:1996vw} loop respectively. An error due to the limited knowledge of
$c_t$ and $\tilde c_t$ is propagated onto our data  and perturbative lattice artifacts are subtracted as detailed
in~\cite{Brida:2016flw}, before trying various fitting procedures.

The left panel of \fig{fig-ssfcont} shows our data and two possible continuum extrapolations, namely
\eq{eq-fit1} with $n_\sigma=3$, $i=2$ and $n_\rho = 2$ or 0.

With the non-perturbatively determined $\beta$ or step-scaling function at hand it is possible
to determine the coupling at renormalization scales 
$L_0^{-1} \to 2L_0^{-1} \to 4L_0^{-1} \ldots$. At step $n$ \eq{eq-Lambda} can be used to obtain
an estimate of $2^{-n} L_0 \Lambda^{(3)}$ with a perturbative truncation error, that decreases with
growing $n$. The situation is depicted in \fig{fig-PTapp}. The final result of this part
is (see~\cite{Brida:2016flw} for further details)
\begin{equation}
   L_{\text 0}\Lambda_{\text{SF}}^{(3)} = 0.0303(8)\, ,\qquad \text{or}\qquad
   L_0 \Lms                             = 0.0791(21) \, .
\end{equation}
The well known relation~\cite{Sint:1995ch} between the SF and the $\overline{\text{MS}}$ schemes
$\Lambda_{\text{SF}}^{(3)} = 0.38286(2)\Lms$
is used here.

\subsection{Gradient Flow Coupling}\label{sec-GF}
A recently significantly 
improved~\cite{Narayanan:2006rf,Luscher:2010iy,Luscher:2011bx,Luscher:2013cpa,Shindler:2013bia,DelDebbio:2013zaa,Fodor:2014cpa,Harlander:2016vzb}
understanding of the Yang-Mills gradient flow (GF)~\cite{Atiyah:1982fa}
has opened the way for the definition of novel scales like $t_0$~\cite{Luscher:2010iy} or $w_0$~\cite{Borsanyi:2012zs} and
for the introduction of new renormalization schemes~\cite{Fodor:2012td,Fritzsch:2013je,Luscher:2014kea}.

Correlation functions of fields $B_\mu$, that are solutions of the gradient flow
equation
\begin{eqnarray*}
  \partial_t B_\mu(t,x) &=& D_\nu G_{\nu\mu}(t,x),\qquad\qquad B_\mu(0,x) = A_\mu(x)\, , \\
  D_\mu                 &=& \partial_\mu + [B_\mu, \cdot]\, ,\\
  G_{\mu\nu}            &=&  \partial_\mu B_\nu - \partial_\nu B_\mu + [B_\mu,B_\nu]\, ,
\end{eqnarray*}
were found to be automatically renormalized at flow times $t>0$.
Simple gauge-invariant combinations, like the action density, can be used in the definitions of
couplings and scales, which can then be measured extremely precisely. Moreover their 
variances are renormalized quantities themselves, such that the continuum limit can be approached
without the problem of a diverging noise to signal ratio.

Already in~\cite{Luscher:2010iy} the proposal was made to define a renormalized coupling based on the dimensionless
combination $t^2\langle G_{\mu\nu}^a G_{\mu\nu}^a\rangle$, where the renormalization scale is
given by the flow time $\mu = 1/\sqrt{8t}$. This original gradient flow renormalization scheme 
has been recently studied to 2 loops in perturbation theory~\cite{Harlander:2016vzb}. 
Different finite volume renormalization 
schemes, where the flow time (and therefore renormalization scale) are tied to the 
box size $\mu = 1/L = c/\sqrt{8t}$, were invented~\cite{Fodor:2012td,Fritzsch:2013je,Ramos:2014kla} and
successfully applied~\cite{Hasenfratz:2014rna,Fodor:2015baa,Rantaharju:2015cne,Fodor:2016zil,Leino:2017lpc}.
The variant that is used in this work
follows~\cite{Fritzsch:2013je}. The running coupling is defined to be
\begin{equation}\label{eq-gGF}
   \ggGF(\mu) = \mathcal{N}^{-1} \frac{t^2}{4} \frac{\left\langle G_{\mu\nu}^a(t,x) G_{\mu\nu}^a(t,x) \, \delta_{Q,0}\right\rangle}
                                                               {\langle \delta_{Q,0}\rangle} \Biggr|_{\sqrt{8t}=cL, x_0=T/2}\, ,
\end{equation}
where $L=T$ is the size of a massless Schr\"odinger functional without background field ($C=C'=0$), $c=0.3$,
$\mathcal{N}$ a computable normalization factor
and the summation over $\mu$ and $\nu$ is restricted to the spacial components only.
$Q = \frac{1}{32\pi^2}\int \! d^4 x\, \epsilon_{\mu\nu\rho\sigma} G_{\mu\nu}^a(t,x) G_{\rho\sigma}^a(t,x)$ is
the topological charge and a projection onto the trivial sector is included in the definition \eq{eq-gGF}.
This projection reduces the variance and simplifies the error analysis in cases where topological 
sectors are sampled  very slowly~\cite{Luscher:2011kk}. Moreover algorithms can be used, that deliberately stay in the 
trivial sector~\cite{DallaBrida:2016kgh}.

On the lattice a discretization has to be chosen for the action, for the flow equation and for the definition 
of the observable $G_{\mu\nu}^a(t,x) G_{\mu\nu}^a(t,x)$. In order to be able to determine the largest simulated box size $\Lhad$ in fm, 
defined implicitly by
\begin{equation}
   \ggGF(\Lhad^{-1}) \equiv 11.31\, , 
\end{equation}
the discretization of the action needs to be the same as the one used by CLS, for which the scale was set.
That means, a tree level Symanzik improved gauge action ($S_{\rm LW}$) and massless clover improved Wilson fermions. In a finite 
volume it becomes necessary to specify, how exactly the SF boundary conditions are 
realized and which values for the boundary improvement coefficients $c_t$ and $\tilde c_t$ are used.
We opt for choice B of ref.~\cite{Aoki:1998qd} for which the coefficients are known to one 
loop~\cite{Takeda:2003he,Pol}.

For the discretization of the flow equation and the observable we follow~\cite{Ramos:2015baa} and use
the Symazik $O(a^2)$ improved lattice flow equation, a.k.a ``Zeuthen flow''
\begin{equation}
   a^2\left[\partial_t V_\mu(t,x) \right]V_\mu(t,x)^\dagger = - g_0^2\left(1+\frac{a^2}{12}\Delta_\mu \right)\partial_{x,\mu}S_{\rm LW}[V],
   \qquad V_\mu(0,x)=U_\mu(x)\, .
\end{equation}
Here $\partial_{x,\mu} S_{\rm LW}$ is the force derived from the improved action.
For the observable we use the (at finite $t$) $O(a^2)$ improved choice that
also enters $S_{\rm LW}$.

\begin{figure}[thb]
  \centering
  \begin{center}
   \includegraphics[width=0.49\linewidth]{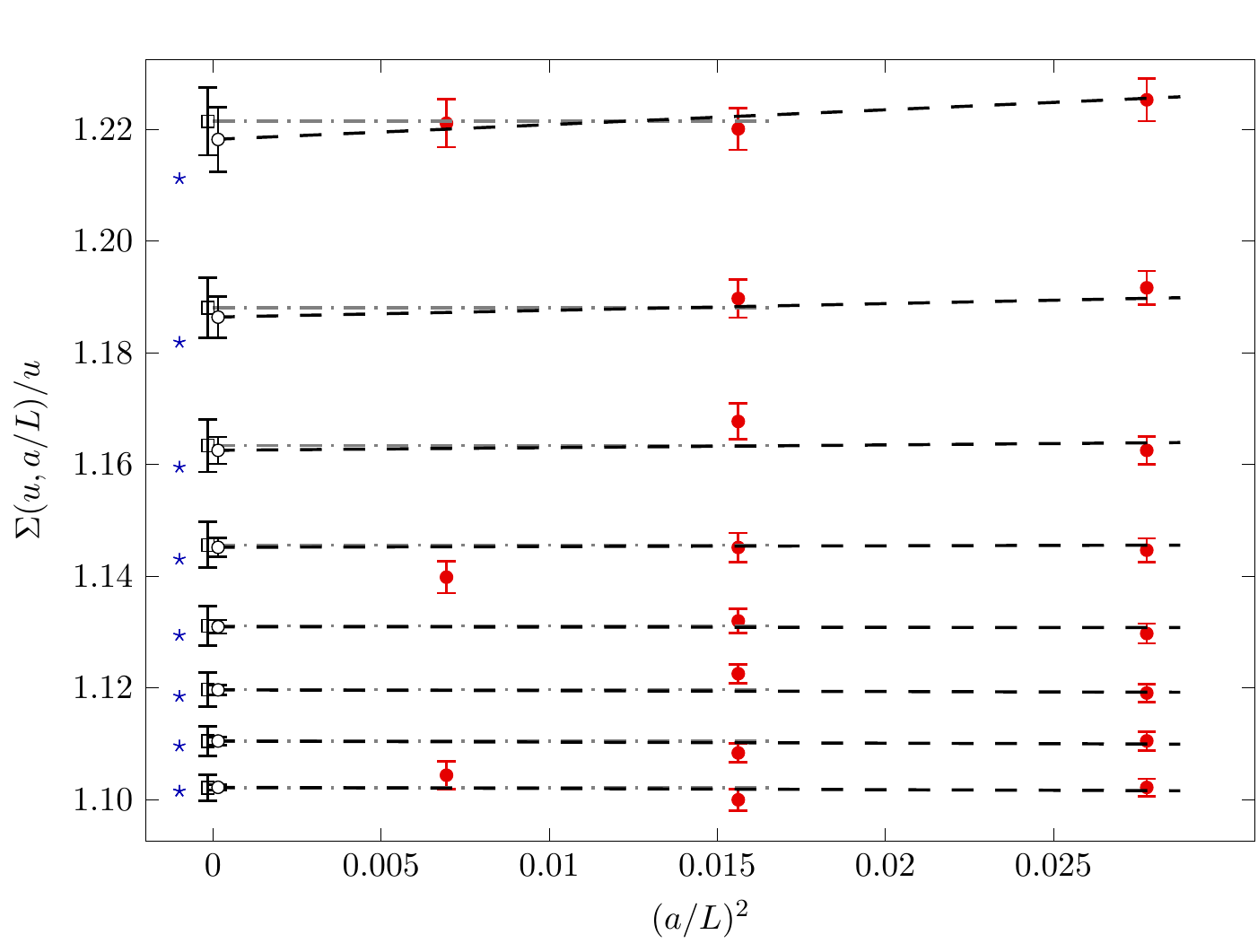}
   \hfill \includegraphics[width=0.49\linewidth]{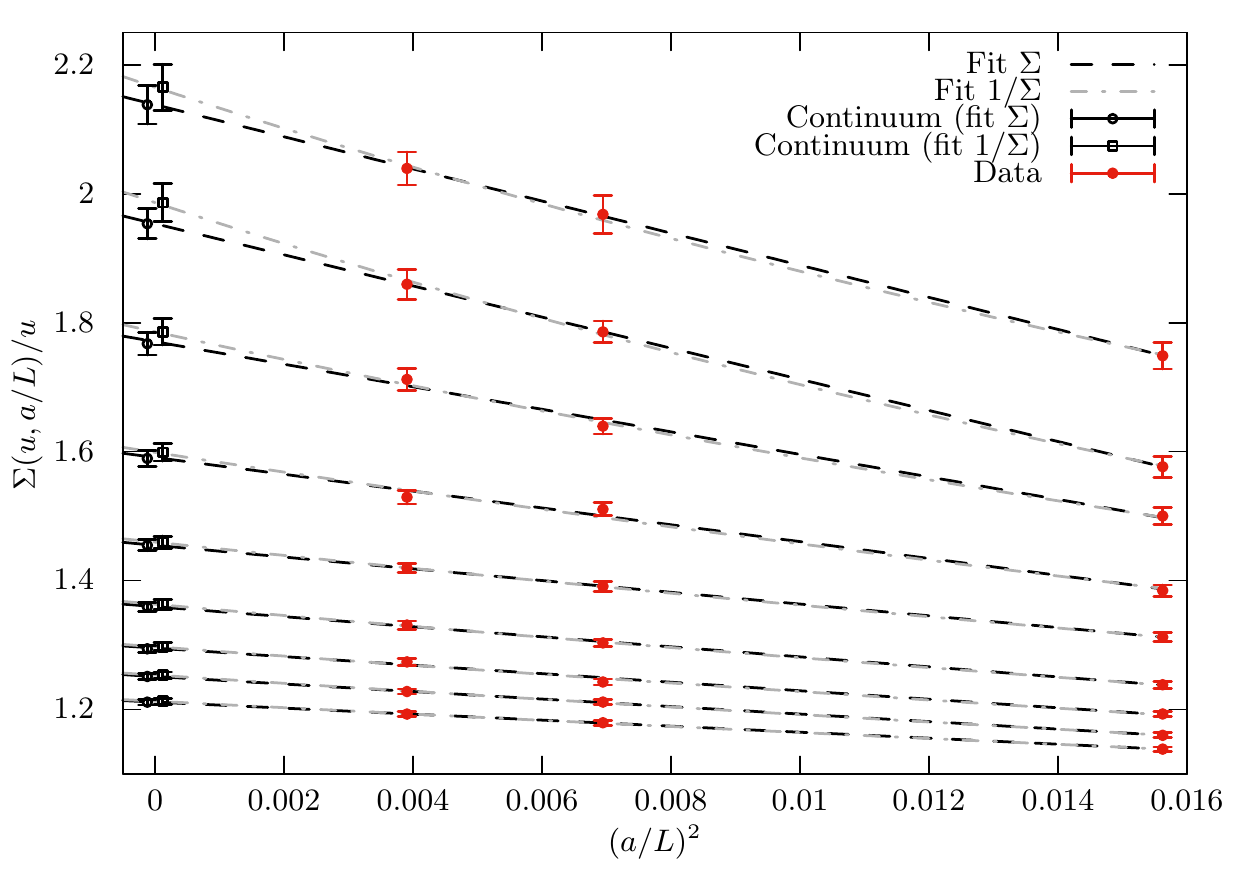}
  \end{center}
  \caption{Exemplary continuum limits of the step-scaling functions in the SF (left) and the GF (right) schemes.
           Asterisks in the left plot mark perturbative predictions.}
  \label{fig-ssfcont}
\end{figure}

Despite systematically removing several sources of $O(a^2)$ lattice artifacts, 
the remaining scaling violations are quite 
significant, especially when compared to those encountered in the SF scheme. 
Consequently, finer lattices were necessary for a controlled continuum extrapolation.
The data set for this part of the project consists of lattices with
$L/a \in \{8, 12, 16 \}$ (and doubled lattices) at couplings
$u\in\{ 2.12, 2.39, 2.74, 3.20, 3.86, 4.48, 5.30, 5.87, 6.55\}$ (approximately).
The right panel of \fig{fig-ssfcont} shows our data and two possible continuum extrapolations.

\begin{figure}[thb]
  \centering
  \begin{center}
   \includegraphics[width=0.5\linewidth]{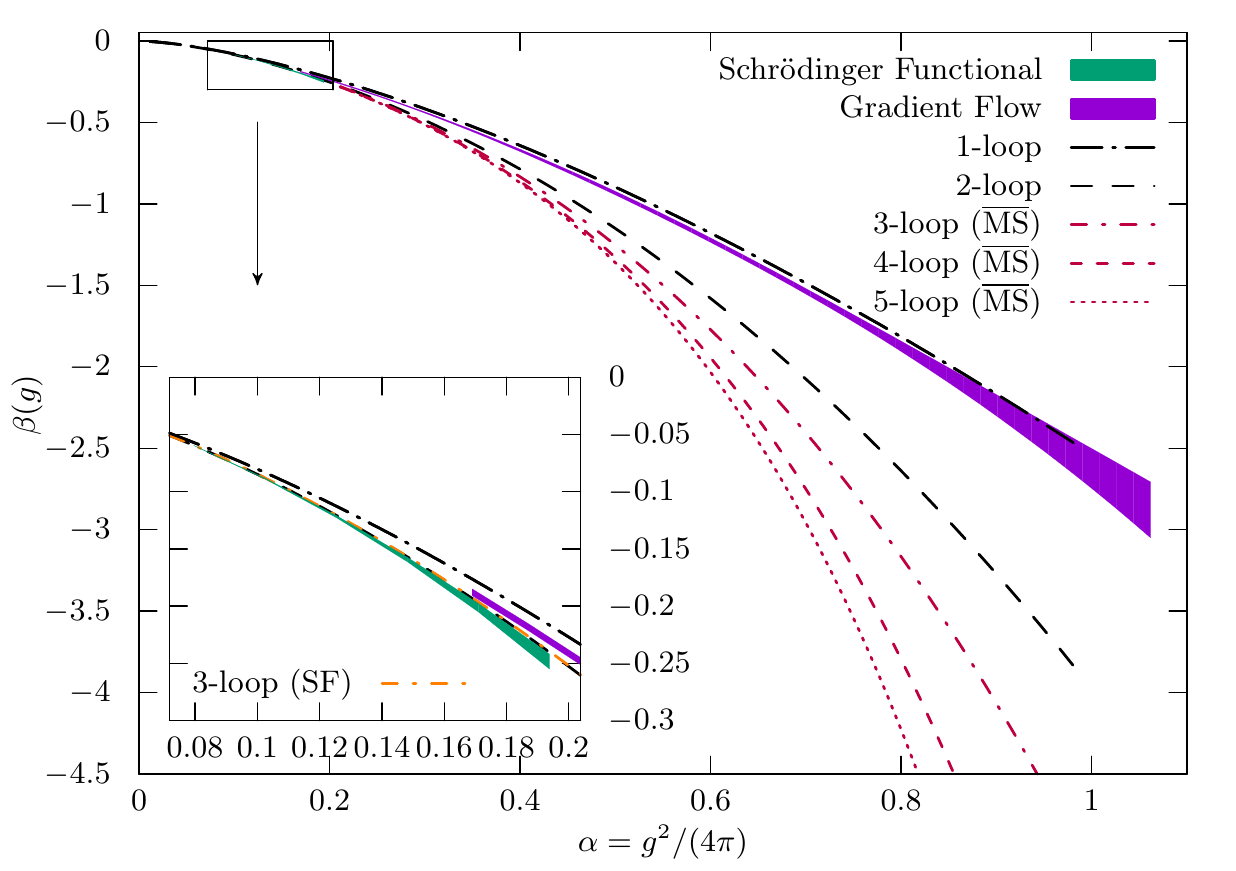}
  \end{center}
  \caption{Non-perturbatively determined $\beta$-functions in the GF and SF schemes and a comparison to various
  perturbative orders.}
  \label{fig-betaall}
\end{figure}

\fig{fig-betaall} shows the non-perturbative continuum $\beta$ functions in 
both schemes in the range of couplings where they were determined.
As expected, in the high energy regime the SF $\beta$-function follows
the perturbative prediction quite closely. In the GF scheme, in the strong 
coupling region, the curve deviates noticeably from the two loop prediction. 
The running is slower, better (but not quite) described by the one loop 
expression.

\subsection{Non-perturbative Matching}\label{sec-Lswi}
\begin{figure}[thb]
  \parbox{.49\linewidth}{
  \centering
  \begin{center}
   \includegraphics[width=\linewidth]{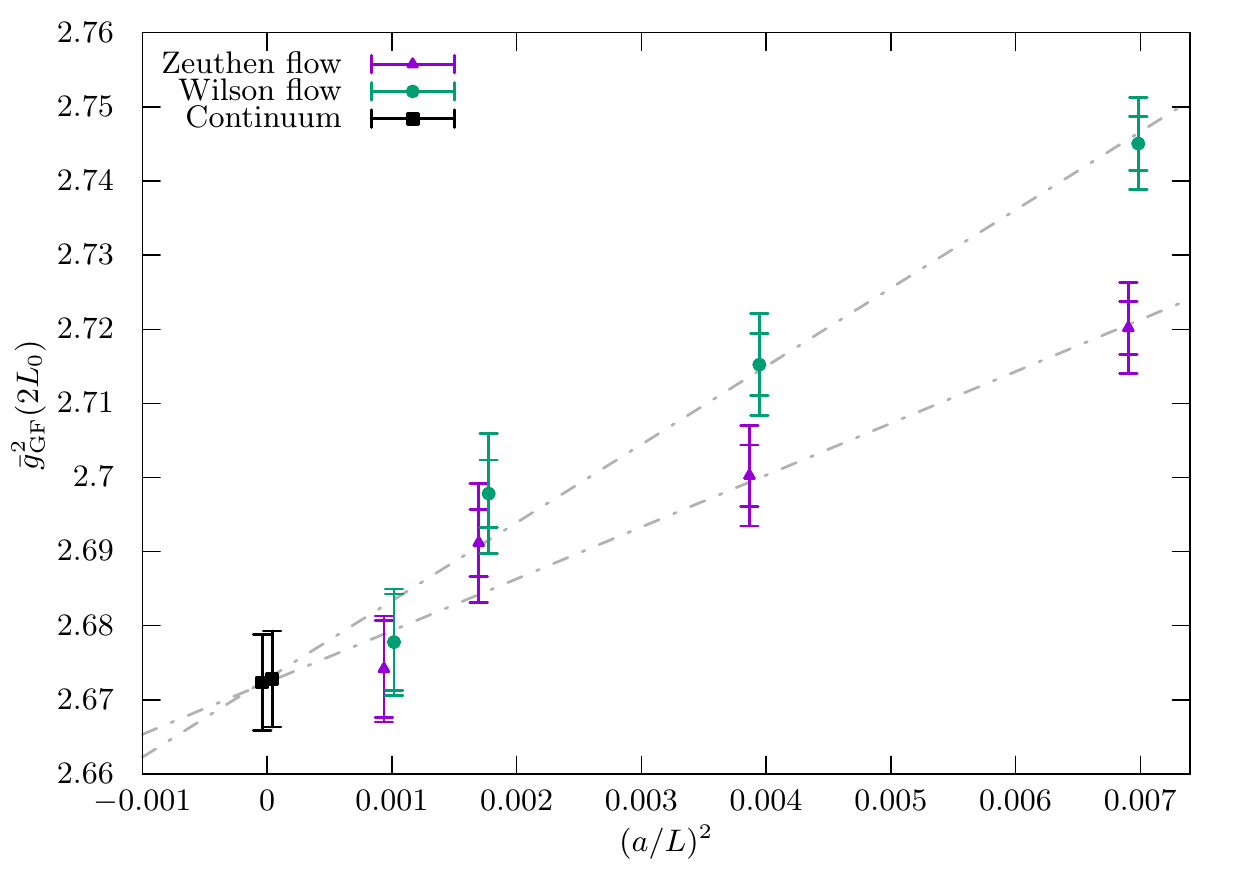}
  \end{center}
  \caption{Non-perturbative matching of SF and GF couplings. Continuum extrapolations are shown
  for the standard (improved) definition of the GF coupling and for one in which the unimproved
  Wilson flow and an unimproved definition of the observable were used. The smaller error bars
  stem from $\ggGF$, the complete error bar contains also the contribution due to a finite precision 
  in the tuning to $\ggSF=2.012$.}
  \label{fig-Lsw}
  }\hfill\parbox{0.49\linewidth}{
  \centering
  \vspace{-1.2cm}
  \begin{center}
   \includegraphics[width=\linewidth]{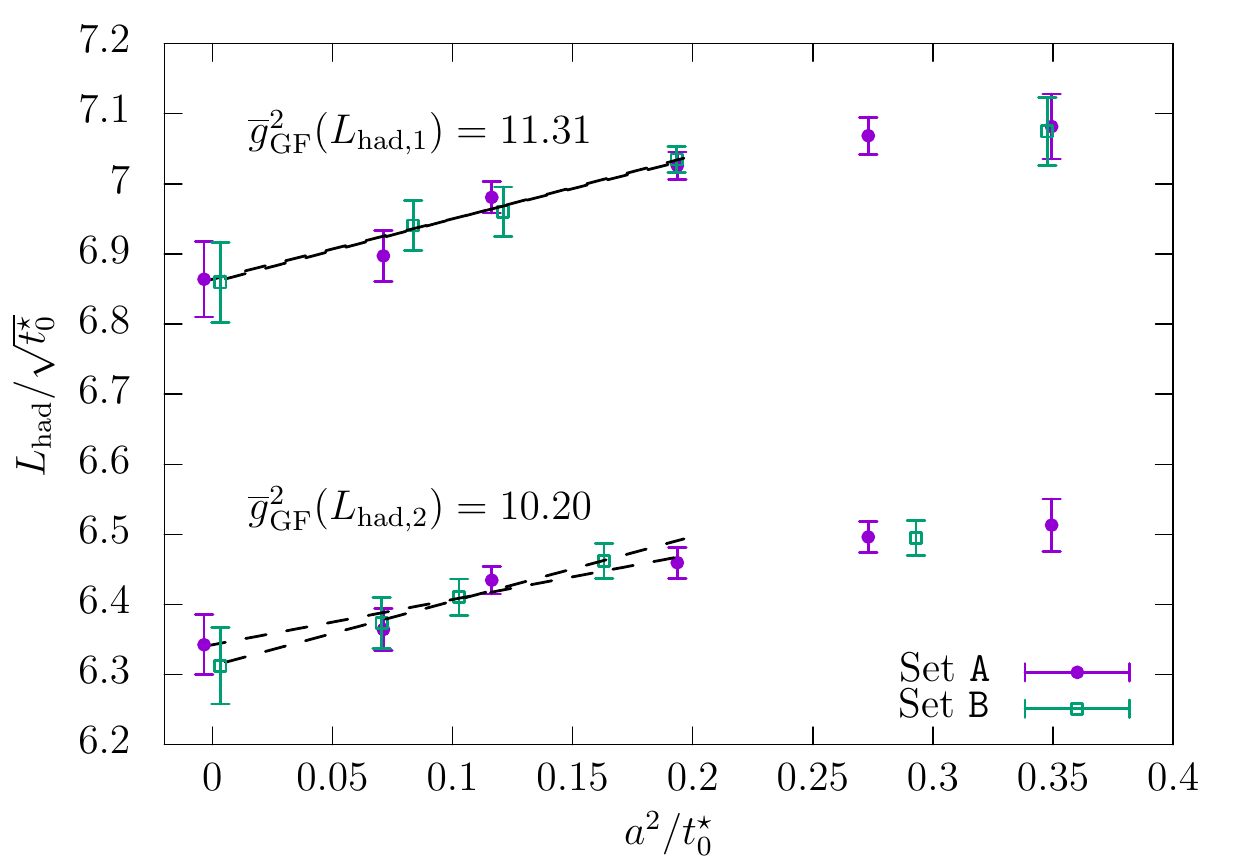}
  \end{center}
  \caption{Determination of the continuum limit of the ratio $\Lhad / \sqrt{t_0^\star}$ for two definitions
  of $\Lhad$. In Set A the data in \tab{tab-Lhad} are interpolated to the improved bare couplings corresponding to \tab{tab-t0}.
  In Set B the other way around.}
  \label{fig-Lhadt0}
  }
\end{figure}

The $\beta$ function of the GF scheme can be used to compute
$L_0$ in fm, if $\Lhad$ in fm is known. To do this, it is
necessary to know the value of the coupling $\ggGF(L_0^{-1})$.
This is not entirely trivial, because $L_0$ is defined implicitly 
in \eq{eq-L0def}, i.e. using a different scheme with different 
boundary conditions and a different lattice action. Since 
$\ggGF$ and $\ggSF$ cannot be measured on the same ensembles a (small) set
of new simulations is necessary. The scheme switch is combined with one
iteration of step-scaling. For every $L/a \in \{12, 16, 24, 32\}$ with
$\ggSF=2.012$ a simulation with the same action, same bare parameters,
doubled linear lattice size and
switched off boundary field is carried out. On these doubled lattices
$\ggGF((2L_0)^{-1})$ is measured. Finally its continuum limit is
obtained, as shown in \fig{fig-Lsw}. Its value, independent of the lattice action, is
\begin{equation}
   \ggGF((2L_0)^{-1}) = 2.6723(64)\, .
\end{equation}
Inserting the numerically determined $\beta$-function into
\begin{equation}
   \ln\left[\frac{2L_0}{\Lhad} \right] = \int\limits_{\bar g_{\rm GF}((2L_0)^{-1})}^{\bar g_{\rm GF}(\Lhad^{-1})} \frac{dx}{\beta(x)}
\end{equation}
yields~\cite{DallaBrida:2016kgh}
\begin{equation}
   \Lhad / L_0 = 21.86(42)\, .
\end{equation}

\subsection{Connection to Infinite Volume}\label{sec-Linf}
To combine everything into a final result according to \eq{eq-Lambdastrat}, the last missing factor 
is $\frac{\sqrt{t_0}}{L_{\text{had}}}$ in the continuum limit.
At finite lattice spacing the numerator is known in lattice units from \tab{tab-t0}. The denominator
is known in lattice units as well - the bare couplings that result in 
$\ggGF = 11.31$ were found in  subsection~\ref{sec-GF} for $\Lhad/a \in \{8, 12, 16\}$. These, and
in addition values of the bare coupling for $\Lhad/a \in \{20, 24, 32\}$ are
summarized in \tab{tab-Lhad}. The tuning has a finite precision and the associated error is propagated onto
the listed $\Lhad/a$ values. 

The action for large volume and GF simulations is, apart from mass terms, the same, but the bare couplings 
in \tab{tab-t0} differ from those in \tab{tab-Lhad}. In order for the lattice spacing to drop out 
in the ratio, the data in \tab{tab-t0} needs to be interpolated to the bare
couplings of \tab{tab-Lhad} or vice versa. Same $g_0$ means same $a$ up to $O(a)$ artifacts. These
can be reduced to $O(a^2)$ if the bare couplings in \tab{tab-t0} are replaced 
by~\cite{Luscher:1996sc,Bhattacharya:2005rb}
\begin{equation}
   \tilde g_0^2 = g_0^2\left(1+\frac{1}{3}{\rm tr}[aM_q]b_g(g_0) \right)\, .
\end{equation}
$M_q$ is the subtracted quark mass matrix and $b_g$ an improvement coefficient, currently known 
only perturbatively~\cite{Sint:1995ch} to one loop.

The interpolations are short and relatively simple. A polynomial 
in $6 / \tilde g_0^2$ is used as an ansatz to fit $\ln(\sqrt{a^2/t_0})$
or $\ln(a/\Lhad)$. Details of the procedure can be found in the supplementary material 
of~\cite{Bruno:2017gxd}. Once the interpolations are done, the continuum limit
of the ratio can be taken as shown in \fig{fig-Lhadt0}. In addition to the standard
definition of $\Lhad$ used so far, a slightly different choice corresponding to
\begin{equation}
    \ggGF(L_\text{had,2}^{-1}) = 10.2
\end{equation}
is shown in the figure. The complete analysis was carried out with both choices, in order to 
assess systematic effects induced by the interpolations.

\eq{eq-Lambdastrat} can now be evaluated. The result is
\begin{equation}\label{eq-Lms3}
   \Lms = 341(12)\ \text{MeV}\, .
\end{equation}

\section{Perturbative Decoupling}\label{sec-dec}
\eq{eq-Lms3} is our main result. It is almost fully non-perturbative. Perturbation theory is
only used at very small couplings $\alpha(\mu) \approx 0.1$ or $\mu \approx 70$ GeV, where it can be
trusted and was tested (\fig{fig-PTapp}).

For a comparison with experimental and other lattice determinations the $\Lambda$-parameter in 
theories with four, five or even six flavors is necessary. We obtain those values based on 
perturbative decoupling.

QCD with $\Nf$ massless quarks and one massive quark, with renormalized mass $\bar m$, 
can be described by an
effective $\Nf$-flavor theory~\cite{Weinberg:1951ss}. To leading order this effective theory is given
by $\Nf$-flavor QCD. Subleading terms give rise to power corrections starting at 
$O(\Lambda^2/\bar m^2, E^2/\bar m^2)$. Up to such corrections, dimensionfull low energy ($E$) 
quantities computed in the effective theory equal those in the $\Nf+1$ theory, if 
the coupling is matched
\begin{equation}
   \bar g^{(\Nf)}(\mu) = \bar g^{(\Nf+1)}(\mu) \times \xi(\bar g^{(\Nf+1)}(\mu),\bar m/\mu) \, .
\end{equation}
Through the matching $\bar g^{(\Nf)}(\mu)$ depends implicitly on $\bar m$. The matching function
$\xi$ is known perturbatively in the $\overline{\text{MS}}$-scheme to four 
loops~\cite{Weinberg:1980wa,Bernreuther:1981sg,Grozin:2011nk,Chetyrkin:2005ia,Schroder:2005hy,Kniehl:2006bg}.
The expansion assumes a particularly simple form if the renormalization scale is chosen
to coincide with the scale $m^*$, defined such, that the running mass at scale $m^*$ equals
$m^*$, i.e. $m^* = \bar m_{\overline{\text{MS}}}(m^*)$. With this choice
the coefficients in
\begin{equation}
    \xi(\bar g,1) = 1 + c_2 \bar g^4 + c_3\bar g^6 + c_4\bar g^8 + O(\bar g^{10})\, 
\end{equation}
are pure numbers and $c_1$ is absent. Together with the perturbative $\beta$ function,
this relation between couplings implies a relation between $\Lambda^{(\Nf)}_{\overline{\text MS}}$
and $\Lambda^{(\Nf+1)}_{\overline{\text MS}}$.
Using as inputs $\Lms$ and $m^*_\text{charm}=1.280(25)$~GeV, $m^*_\text{bottom}=4.180(30)$~GeV, 
\cite{Patrignani:2016xqp}
and $m^*_\text{top} = 165.9(2.2)$~GeV~\cite{Fuster:2017rev} the four, five and six 
flavor $\Lambda$ parameters can be computed.

Two questions arise. How large are the neglected power corrections? Can perturbation
theory be trusted at $\mu=m^*_\text{charm}$?

The first question has been addressed non-perturbatively in a simplified setup, where the 
decoupling was investigated for QCD with just two heavy 
quarks~\cite{Bruno:2014ufa,Knechtli:2017xgy,thiscontrib24}. At the charm quark mass the 
power corrections in the investigated quantities were found to be tiny, 
typically around two permille. It is more difficult to give a definitive answer to the 
second question. It was addressed in~\cite{Knechtli:2015lux} with encouraging results,
i.e. that PT might be applicable for decoupling at the charm scale.

Here we can only test decoupling within perturbation theory itself. To do 
so, we translate our result for $\Lms$ to $\Lambda_{\overline{\text{MS}}}^{(5)}$ and then,
inverting \eq{eq-Lambda}, to $\alpha_{\overline{\text{MS}}}^{(5)}(M_\text{Z})$, with
$M_\text{Z}=91.1876$, \cite{Patrignani:2016xqp}. This is done in perturbation theory with $n$ loop accuracy
for the decoupling relations and $n+1$ loop accuracy for the $\beta$ function. The apparent
convergence of the final result with $n$ is monitored and turns out to be excellent.
The sequence is tabulated in \tab{tab-dec}. As a systematic error the difference
between the $n=4$ and the $n=2$ result is taken, which, within perturbation theory, is conservative.
It plays a minor role in the overall error budget.

\begin{table}[thb]
  \small
  \centering
  \caption{PT decoupling}\label{tab-dec}
  \begin{tabular}{c c c}
         \toprule
         n (loops) & $\alpha^{(N_\text{f}=5)}_{\overline{\text{MS}}}$ & $ \alpha_n-\alpha_{n-1} $\\
         \midrule
         1         & 0.11699                                        &                        \\
         2         & 0.11827                                        & 0.00128                \\
         3         & 0.11846                                        & 0.00019                \\
         4         & 0.11852                                        & 0.00006                \\
         \bottomrule
  \end{tabular}
\end{table}

\section{Conclusions}\label{sec-res}
Our final results are
\begin{eqnarray}
   \Lms                                            &=& 341(12)\ \text{MeV}\, , \\
   \Lambda_{\overline{\text{MS}}}^{(4)}            &=& 298(12)(03)\ \text{MeV} \ \ \ \qquad \text{(pert.\ decoupling)}\, ,\\
   \Lambda_{\overline{\text{MS}}}^{(5)}            &=& 215(10)(03)\ \text{MeV} \ \ \ \qquad \text{(pert.\ decoupling)}\, ,\\
   \Lambda_{\overline{\text{MS}}}^{(6)}            &=& 91.1(4.5)(1.3)\ \text{MeV} \qquad \text{(pert.\ decoupling)}\, ,\\
   \rightarrow \alpha^{(5)}_{\overline{\text{MS}}}(m_\text{Z}) &=& 0.1185(8)(3)\, ,\\
                                                   & & 0.1174(16)\hspace{2cm} \text{PDG\ non-lattice}\, .
\end{eqnarray}
We reach a precision in $\alpha_\text{S}$ of slightly below $0.7\%$ which  is very good - twice as precise as the non-lattice 
world average. The main sources of errors are the statistical errors of the step-scaling functions. They contribute
$22\%$ (GF) and $50\%$ (SF) to the total relative error squared of $\alpha_\text{S}$. The introduction of the second finite volume scheme, 
based on the gradient flow has paid off. The usually most problematic region of large couplings does not dominate
the error as it did in the past. Observing the relatively large contribution to the error stemming from 
non-perturbative evolution of the SF coupling between 5 GeV and 70 GeV, the question arises whether one maybe could have
used perturbation theory already at lower energies than we did. 
\begin{figure}[thb]
  \centering
  \begin{center}
   \includegraphics[width=0.5\linewidth]{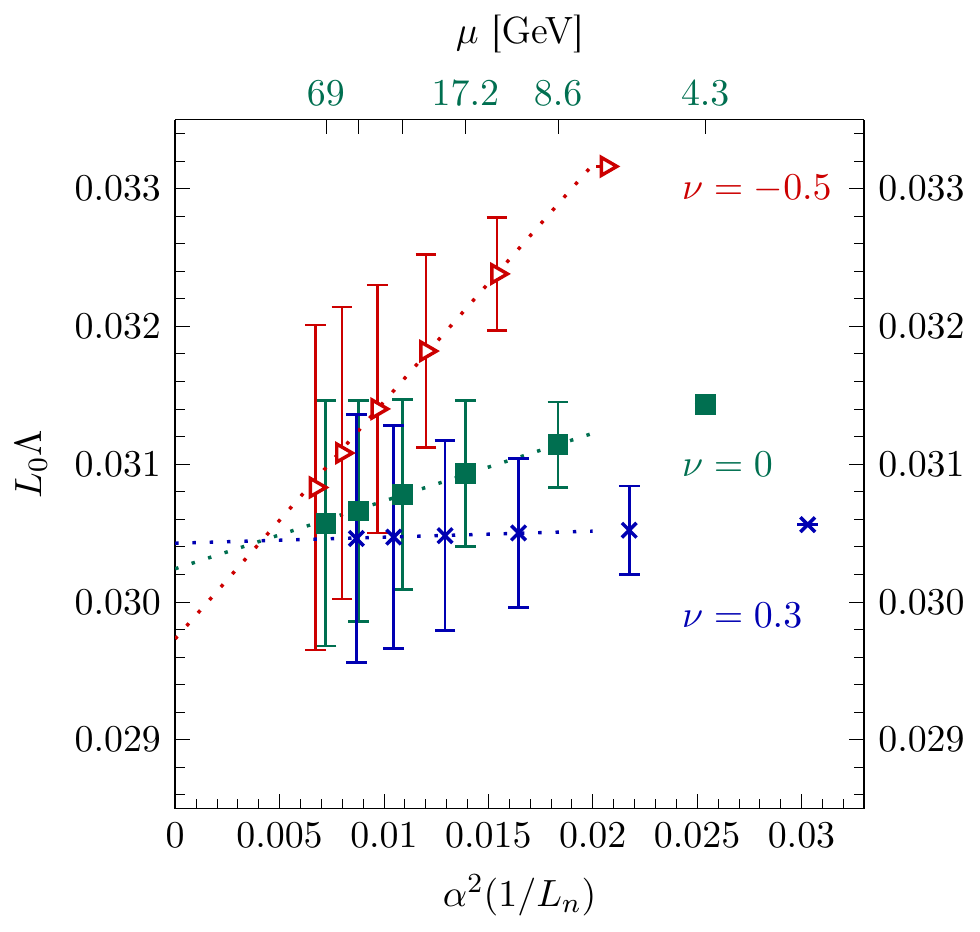}
  \end{center}
  \caption{Results for $\Lambda$ parameters in different SF schemes, depending on  the energy from which on three-loop perturbation theory 
           was used.}
  \label{fig-PTapp}
\end{figure}
\fig{fig-PTapp} shows the $\Lambda$ parameter in units of $L_0$, that one would obtain
by using the perturbative 3-loop $\beta$-function starting from different energies.
Purely perturbative behavior would show as a plateau in this plot. 
Although results starting from  $\alpha \lesssim 0.14$ are compatible with each other, we
see a drift of the central value. This linear behavior of $L_0\Lambda$
as a function of $\alpha^2$ can be explained by the presence of an effective $b_3$ term in the 
$\beta$-function. Relying on the 3-loop formula already at $\alpha \approx 0.14$ would reduce the 
statistical error significantly, but also introduce a systematic error of the order of $1\sigma.$
Such statements are highly scheme dependent. For instance \fig{fig-PTapp} shows also the 
situation for two slightly different SF-couplings -- the truncation error can become much
worse, or completely absent, depending on the chosen scheme. The family of SF schemes considered
here is generally believed to be particularly well behaved perturbatively. In other
schemes truncation error could be much worse. An example is the $\alpha_{qq}$ scheme. Although its
$\beta$-function is known to four loops, a significant deviation between perturbative and
non-perturbative running can be observed at $\alpha_{qq} \approx 0.24$,~\cite{thiscontrib469}.

For any future significant (e.g. factor $1/2$ in the error) improvements of our programme 
several challenging problems have to be overcome. 
The limited knowledge of boundary improvement coefficients will
make pure $O(a^2)$ continuum extrapolations difficult. The subleading
cutoff effects in $\ggGF$ will become significant and will have to be addressed, for instance by
considering even finer lattices than the $16^4 \to 32^4$ ones used here. Electromagnetic and
iso-spin effects in all quantities used for scale-setting will have to be accounted for.
Finally, a fourth dynamical quark should better be considered in order to shift the scale at 
which one relies on perturbative
decoupling up to the bottom mass.


\section*{Acknowledgments}
I would like to thank my colleagues in the ALPHA collaboration and in the CLS
consortium for a pleasant and fruitful collaboration, in particular my co-authors
in~\cite{Bruno:2014jqa,Bruno:2016plf,Brida:2016flw,DallaBrida:2016kgh,Bruno:2017gxd}.
Francesco Knechtli and Rainer Sommer provided valuable comments on the manuscript for which
I am deeply grateful.

We thank Pol Vilaseca for sharing his result on $\tilde c_t$ with us and
Carlos Pena, Isabel Campos and David Preti for helping us generate
several of the ensembles for the GF coupling.

The generation of large volume CLS ensembles would not have been possible without
access to major computational resources.
We acknowledge PRACE for awarding us access to the resource FERMI at CINECA and SuperMUC at the 
LRZ. We thank the Swiss National Supercomputing Centre for supporting us by a grant 
under project ID s384. We gratefully acknowledge the Gauss Centre for Supercomputing (GCS) for 
providing computing time through the John von Neumann Institute for Computing (NIC) on the GCS share of 
the supercomputer JUQUEEN at J\"ulich Supercomputing Centre (JSC). 
We thank the Helmhotz Institute Mainz and the University of Mainz for 
access to the ``Clover'' HPC Cluster.
The generation of the finite volume ensembles required significant resources as well. We were granted computer time
and support by HLRN (bep0040), by NIC at DESY, Zeuthen and on Altamira at IFCA for which we are grateful.

This work is based on previous work~\cite{Sommer:2015kza} supported strongly by the Deutsche Forschungsgemeinschaft in the 
SFB/TR 09.

\bibliography{lattice2017}

\end{document}